\documentclass[journal]{IEEEtran}
\ifCLASSINFOpdf
   \usepackage[pdftex]{graphicx}
\else
\fi

\usepackage{amsmath,amssymb,amsfonts}
\usepackage{algorithm}
\usepackage{algorithmic}
\usepackage{graphicx}
\usepackage{textcomp}
\usepackage{amsmath}
\usepackage{xcolor}

\IEEEoverridecommandlockouts
\IEEEpubid{\makebox[\columnwidth]{978-1-7281-6680-3/20/\$31.00~\copyright2020 IEEE \hfill} \hspace{\columnsep}\makebox[\columnwidth]{ }}

\begin{document}
\title{Leveraging lightweight blockchain to establish data integrity for surveillance cameras}

\author{\IEEEauthorblockN{Regio A. Michelin\IEEEauthorrefmark{1}\IEEEauthorrefmark{2}, Nadeem Ahmed\IEEEauthorrefmark{1}\IEEEauthorrefmark{2}, 
        Salil S. Kanhere\IEEEauthorrefmark{1}\IEEEauthorrefmark{2},
        Aruna Seneviratne\IEEEauthorrefmark{1}\IEEEauthorrefmark{2} and
        Sanjay Jha\IEEEauthorrefmark{1}\IEEEauthorrefmark{2}}

\IEEEauthorblockA{\IEEEauthorrefmark{1}University of New South Wales (UNSW) - Sydney, Australia}

\IEEEauthorblockA{ \IEEEauthorrefmark{2}Cyber Security CRC - Australia}

\IEEEauthorblockA{E-mail: \{regio.michelin, nadeem.ahmed, salil.kanhere, aruna.seneviratne, sanjay.jha\}@cybersecuritycrc.org.au}}

\maketitle

\IEEEpubidadjcol

\begin{abstract}
The video footage produced by surveillance cameras is important evidence to support criminal investigations. Video evidence can be sourced from public (trusted) as well as private (untrusted) surveillance systems. This raises the issue of establishing integrity for information provided by the untrusted video sources. In this paper, we present a framework to ensure the data integrity of the stored videos, allowing authorities to validate whether video footage has not been tampered with. Our proposal uses a lightweight blockchain technology to store the video metadata as blockchain transactions to support the validation of video integrity. Our evaluations show that the overhead introduced by employing the blockchain to create the transactions introduces a minor latency of a few milliseconds.
\end{abstract}

\begin{IEEEkeywords}
Blockchain, Surveillance Cameras, Integrity.
\end{IEEEkeywords}

\IEEEpeerreviewmaketitle

\section{Introduction} \label{sec:introduction}

\IEEEPARstart{S}{urveillance} cameras are increasingly being used for safety, security, traffic monitoring and law enforcement purposes. The prevalence of these cameras is a result of advances in admissibility of the video footage as criminal evidence in court actions~\cite{Nedim:2019},~\cite{Martin:2019},~\cite{Buchanan:2019}. These cameras are deployed in different places such as homes, shops, malls and offices~\cite{Ashby2017} to inhibit illegal actions. Typically, the video streams of these privately owned cameras are stored privately and only made available to the law enforcement agencies on request. The latter have to rely on watermarking and time stamping provided by the device manufacturer for validating the stored video. There is no guarantee that the obtained video stream has not been digitally tampered with. The variability of these video sources hence raises issues of information trust, authenticity and integrity. This highlights the need for a technology solution that can provide proof of integrity for video surveillance information exchanged between devices operated by entities with different levels of trust. 

Among the new technologies that potentially could address these issues, the blockchain has drawn particular interest as it was initially proposed as a public ledger to maintain Bitcoin~\cite{Nakamot:2009} transactions. However, many changes have since been proposed in the blockchain structure, algorithms and data models to make it suitable for use in different application domains. In the context of video surveillance, we require a lightweight blockchain framework that is suitable for the resource constrained IoT environment and introduces minimal latency in managing transactions. Out of the available IoT based blockchain solutions, we employed a framework called SpeedyChain~\cite{Michelin:2018}, based on its unique capability to allow appending multiple transactions in existing blocks as opposed to traditional blockchains that can only add transactions at block creation time. In SpeedyChain each device has its own block, and all transactions from that device are stored in that block, thus considerably reducing the transaction processing time. This lightweight permissioned blockchain implementation runs at the gateway level and manages transactions received from different sources.


Our specific contributions in this paper are as follows: (\textit{i}) Propose a blockchain based framework to support verifiable video metadata management; (\textit{ii}) System implementation and evaluation using the Raspberry Pi 3 platform; (\textit{iii}) Evaluation of the system's scalability and the overhead introduced.

\section{Proposed Framework} \label{sec:solution}


The proposed framework follows a three-layer architecture presented in Figure~\ref{fig:scenarioDetailed}. The surveillance cameras are assumed trusted and deployed in the sensing layer. The gateways are also trusted and deployed in the transportation layer and are responsible for video streaming, maintaining the blockchain and providing the proof for video integrity. Finally, we have the untrusted third party storage layer where we can use any suitable storage system. For this work, we use Interplanetary File System Network (IPFS) for storing the surveillance videos.


\textit{A. Device bootstrap process}: The bootstrap process takes place when a gateway identifies that there is no existing block in the blockchain containing the camera public key. Each camera is uniquely identified by its public key present in the block header. The block is created and follows the PBFT consensus protocol execution~\cite{lunardi2019} to insert it into the blockchain. Only after the consensus is reached, the block is inserted in the blockchain and the permissioned camera is allowed to start the video streaming.

\textit{B. Video integrity protocol\label{sec:video}}: Figure~\ref{fig:scenarioDetailed} presents the process of creating a transaction of the video feed from surveillance cameras. At the sensing layer, each surveillance camera produces the video streaming which is transferred to the gateway. The gateways are responsible for processing the video stream and forwarding it to the storage system. We explain the functionality of the gateways in several steps:
\begin{figure}[!t]
    \centering
    \includegraphics[width=0.49\textwidth]{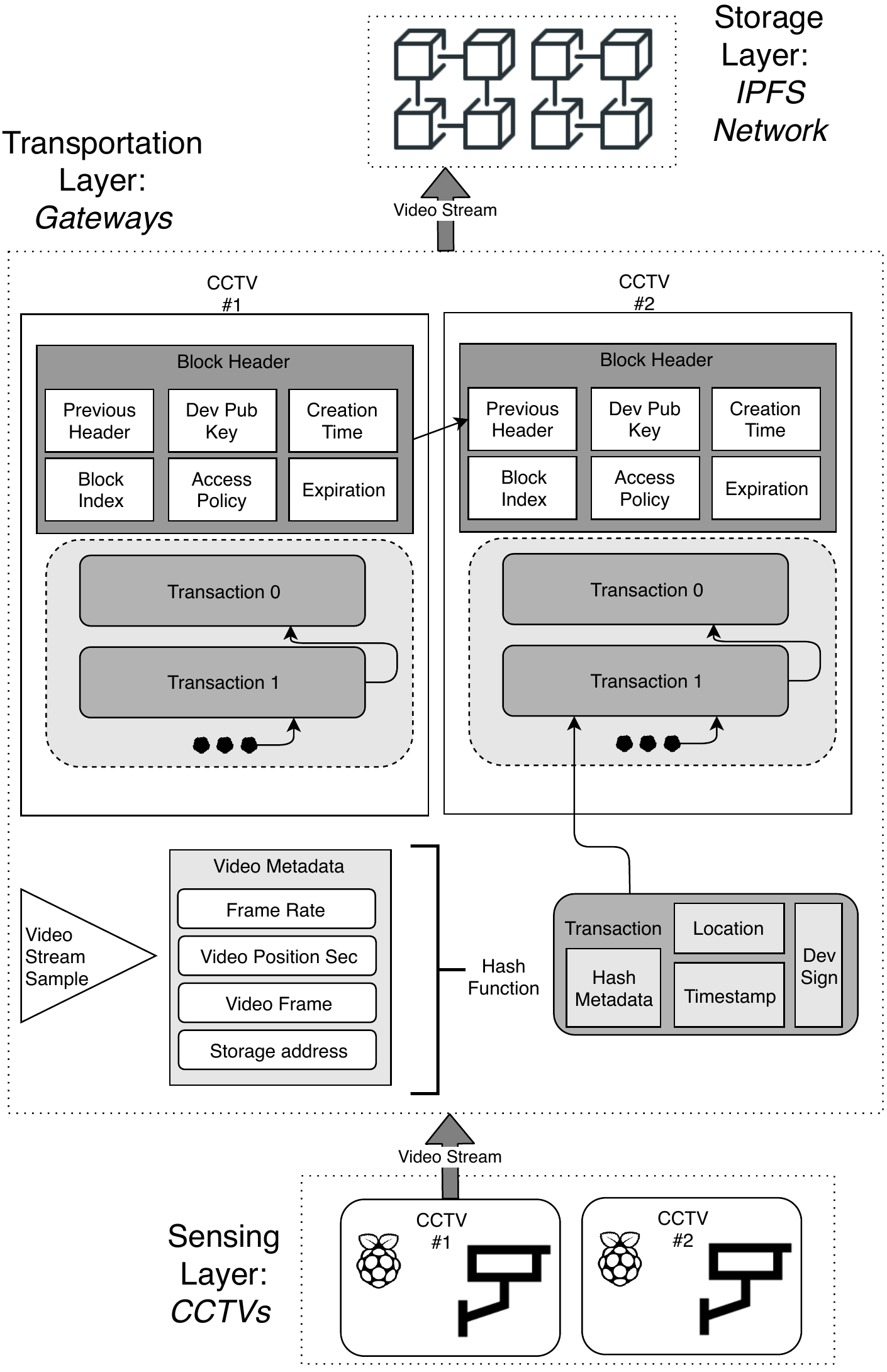}
    \caption{Process for creation of transactions based on video metadata}
    \label{fig:scenarioDetailed}
\end{figure}

\textbf{Step 1)} Once the video chunks are received at the gateway, it extracts the video metadata (\textit{VM}) at regular time intervals ($m$). The metadata is composed of the width (\textit{Wi}), height (\textit{He}), frame rate (\textit{Fr}), current position (\textit{Po}) indicated by time in milliseconds, and video hash (\textit{Vh}) of the chunk of video since last interval. The gateway next computes the metadata hash value ${HashVM_m = Hash(Wi,He,Fr,Po, Vh)}$ which will used for ensuring the integrity of the video. 

\textbf{Step 2)} Once the gateway has calculated the $HashVM_m$, the video chunk is forwarded to the IPFS storage network. Each new chunk that has been pushed into IPFS is accessed by the address that IPFS has generated. This address is required for future access to the video and for validation purposes.

\textbf{Step 3)} The gateway can now proceed with creating a transaction that is composed of; previous hash transaction, sequence number, and the information signed by the gateway. The transaction information field stores the file storage address (obtained in Step 2), the metadata hash (calculated in Step 1), and the timestamp.

\textbf{Step 4)} This transaction is then pushed into the blockchain and a block notification update is published to the peer gateways, to keep the blockchain synchronized.

\begin{figure}[htb]
    \centering
    \includegraphics[width=0.45\textwidth]{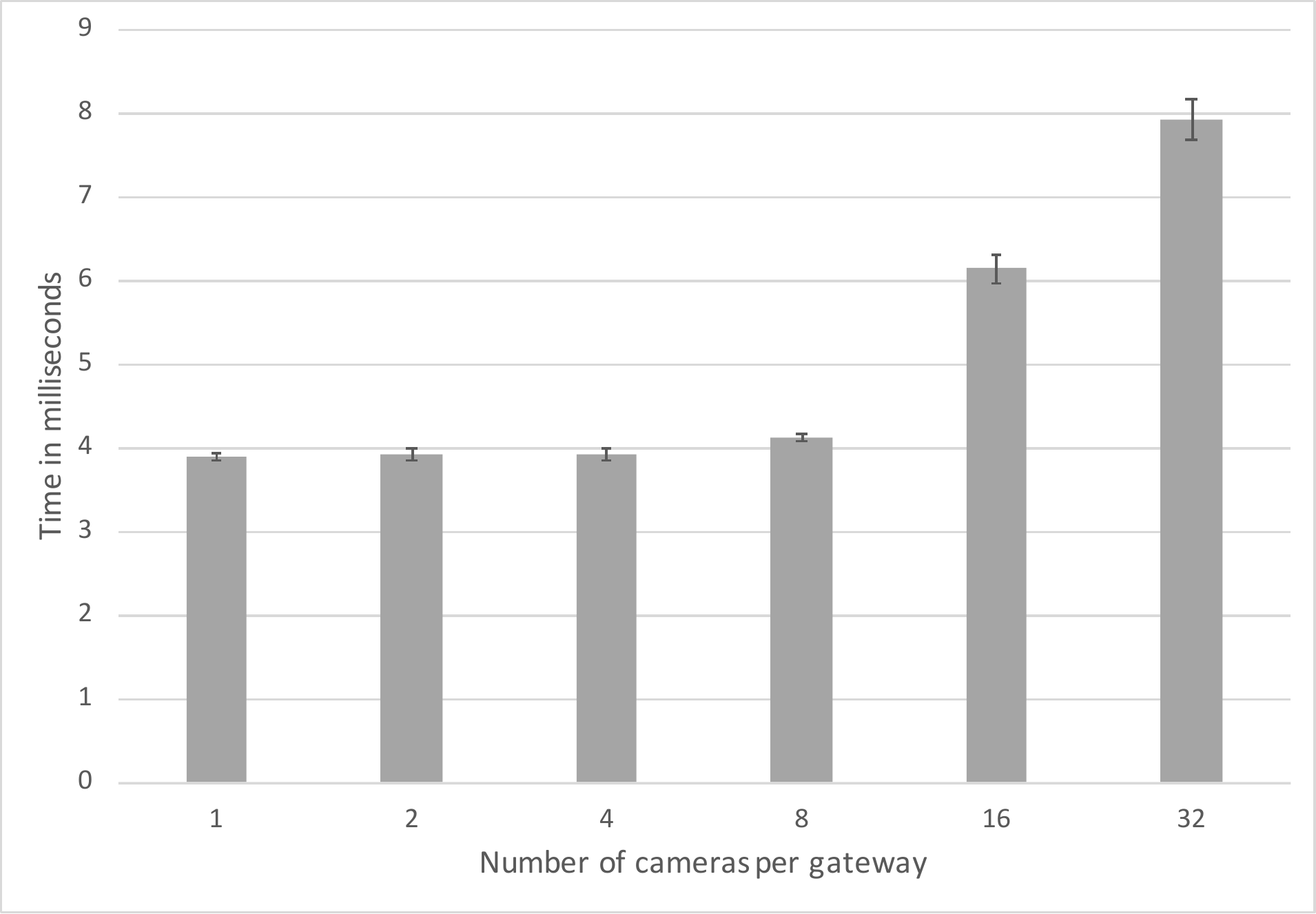}
    \caption{Time to create a video metadata transaction}
    \label{fig:AddValidate}
\end{figure}

\section{Evaluation and Conclusion} \label{sec:evaluation}


The setup uses a video camera module connected to a Raspberry Pi 3 acting as the surveillance cameras. The SpeedyChain and the video streaming functions are deployed in four gateways operating at the transportation layer. The IPFS storage solution was configured to run in a private instance, allowing the local video stream storage.


The experiment aims to evaluate the overhead involved in creating a new transaction containing the video metadata. We scaled the number of cameras managed by each gateway from 1 to 32. The length of the video processed at the gateway level was of 30 minutes, which was split in small video chunks each of duration 10 seconds to generate 180 transactions for each surveillance camera.

Figure~\ref{fig:AddValidate} presents the results plotting the average processing time against varying number of cameras. The y-axis represents the average processing time to create a new transaction from the metadata information extracted from the video chunk and push it into the blockchain. The graph shows that the average processing time is almost constant when the number of cameras are increased to 8 per gateway, and higher processing times are observed when more than 16 cameras are introduced per gateway. The increase in the processing time beyond 8 cameras per gateway can be attributed to the limited hardware resources assigned to each of the gateway.  

The processing time here includes all the 4 steps from video record protocol (Section~\ref{sec:video}-B). However, as compared with a traditional system, the real penalty only involves step 1, where we calculate the hash of the metadata. The stream is immediately pushed to the IPFS in step 2 while steps 3 and 4 can be considered offline in a way that they do not effect the overall latency of a real time monitoring system. Moreover, even in the worst case scenario where we have 32 cameras per gateway and we consider latency of all 4 steps, the total latency introduced is only about 8 milliseconds.

\section*{Acknowledgment}
The work has been supported by the Cyber Security Research Centre Limited (CSCRC) whose activities are partially funded by the Australian Government's Cooperative Research Centres Programme.

\ifCLASSOPTIONcaptionsoff
  \newpage
\fi

\bibliographystyle{IEEEtran}
\bibliography{reference}

\end{document}